\documentclass[showpacs, pra, aps, twocolumn,10pt]{revtex4}
\usepackage{mathrsfs}
\usepackage{array}
\usepackage{amsmath}
\usepackage{graphicx}
\usepackage{enumerate}
\usepackage{amstext}
\usepackage{bm}
\usepackage{amsfonts}
\usepackage{bbm}
\usepackage[utf8]{inputenc}

\begin{document}
\title{Chirality-induced one-way quantum steering  between two waveguide-mediated ferrimagnetic microspheres}
\author{Huiping Zhan$^{1}$}
\author{Lihui Sun$^{2}$}
\author{ Huatang Tan$^{1}$}
\email{tht@mail.ccnu.edu.cn}
\affiliation{$^{1}$Department of Physics, Huazhong Normal University, Wuhan 430079, China\\
$^{2}$Institute of Quantum Optics and Information Photonics, School of Physics and Optoelectronic Engineering,
Yangtze University, Jingzhou 434023, China}

\begin{abstract}
One-way quantum steering is of importance for quantum technologies, such as secure quantum teleportation. In this paper, we study the generation of one-way quantum steering  between two distant yttrium iron garnet (YIG) microspheres in chiral waveguide electromagonics. We consider that the magnon mode with the Kerr nonlinearity in each YIG sphere is chirally coupled to left- and right-propagating guided  photons in the waveguide. We find that quantum steering between the magnon modes is absent with non-chirality but is  present merely in the form of one way (i.e., one-way steering) when the chirality occurs. The maximal achievable steering is obviously improved as the chirality degree increases. We further find that when the waveguide's outputs are subjected to continuous homodyne detection, the steering can be considerably enhanced and asymmetric steering with strong entanglement can also be achieved by tuning the chirality. Our study shows that chirality can be explored to effectively realize one-way quantum steering. Compared to other studies on achieving asymmetric steering via controlling intrinsic dissipation, e.g. cavity loss rates, our scheme merely depends on the chirality enabled via positioning the micromagnets in the waveguide and is continuously adjustable and experimentally more feasible.

\end{abstract}
\maketitle

\section{Introduction}
Nonclassical states of macroscopic objects \cite{FFr} is of importance for testing fundamental principles of quantum mechanics \cite{Ipik,CPfi}, e.g., decoherence effect at large mass scale \cite{WHZ1,WHZ2,ABas}. Recently, the preparation of nonclassical effects in high-quality ferrimagnetic materials, especially  YIG \cite{HC}, has attracted extensive attention, due to high spin density and low loss rate of magnons, i.e., the quanta of collective excitation of spins in YIG samples. Further, magnon exhibits an excellent ability to interact with a variety of systems, such as microwave photons \cite{oo,HH,YT,XZ,MGOR,LBAI,DKZ,JBOU,NKO}, optical photons \cite{AOsa,XZh,JAH}, phonons \cite{XZh1,JLi,MYu,CAPo}, and superconducting qubits \cite{YTab,DLa1,DLa2,SPWo},  which shows that magnons can be a potential candidate for studying quantum effects in macroscopic-size objects.

Quantum steering \cite{HMW,YXi,SHL} is a kind of quantum nonlocality which is intermediate between entanglement \cite{RHo} and Bell nonlocality \cite{Bell}. Distinct from entanglement and Bell nonlocality, steering can be asymmetric and even one-way with respect to two observers involved. One-way steering, which means that one observer can remotely steer the quantum states of the other but not vice versa, is of importance for secure quantum teleportation \cite{MRei,QHe}, one-sided device-independent
quantum cryptography \cite{EPas,PSkr}, and quantum channels discrimination \cite{MPia}.
Theoretical studies have revealed asymmetric steering effect in various systems, such as optomechanical systems \cite{QYH1,SKie,HTan1,HTan2}  and cavity magnonic systems \cite{HTan4,SSZ,ZBY,HTan5,kong}, mainly achieved with unbalanced intrinsic losses. One-way Gaussian steering has been experimentally observed by controlling the unequal dissipation of two entangled beams \cite{VHan}.

Recent studies on chiral quantum optics have attracted a lot of attention, which offer a novel platform for quantum control of light-matter interactions \cite{PLo}.
In the chiral configurations, such as spin-waveguide systems \cite{DE,EV,AGT,AG,AFv,AFK,MM}, the emitter-photon interaction is non-reciprocal, i.e, ``chiral coupling"-- a manifestation of optical spin-orbit coupling \cite{Bli1,Bli2,Aie}. That is, the coupling of emitters to photons in the waveguide depends on the polarization of the emitter's transition dipole moment and the propagation direction of traveling photons. Photon emission with directionality has been experimentally demonstrated in chiral waveguides \cite{HPi,CGon,MSc,PSc,CAD}. The chirality opens up a new means of controlling quantum effects and  becomes a key ingredient for a range of elementary quantum devices based on chiral quantum effects, such as non-reciprocal single-photon devices \cite{Soll,Xia} and non-destructive photon detectors \cite{KKo}.

In this paper, we propose a chiral route to the generation of one-way quantum steering between two YIG spheres in  waveguide electromagonics.  The YIG spheres are placed in special positions in a microwave waveguide and each magnon mode with the Kerr nonlinearity can be chirally coupled to  left- and right-propagating guided photons in this waveguide.  We reveal how the chirality allows  realizing one-way steering of the two magnon modes, which is unachievable in the non-chiral coupling situation.  Moreover, we further find that the steering can be  enhanced significantly  by homodyne detections applied on the outputs of the waveguide, and asymmetric steering with strong entanglement can also be achieved by tuning the chirality in this situation. Finally, to verify and apply the generated steering, state-based feedback is introduced to convert the conditional results into the unconditional ones with high fidelity. Our study shows  the potential of  chirality  for  realizing one-way quantum steering  protocols. Compared to other studies on manipulating asymmetric steering via unbalanced dissipation, our scheme is experimentally more flexible and controllable since merely depends on the chirality enabled via positioning the micromagnets in the waveguide.

This paper is organized as follows. In Sec. II, the chiral magnon-waveguide system is introduced. In Sec. III, the results are presented in detail.  In Sec. V, the indirect feedback is introduced to achieve unconditional entanglement and steering.  In the last section, some discussion
and the conclusion are given.

\begin{figure}[t]
\centerline{\scalebox{0.52}{\includegraphics{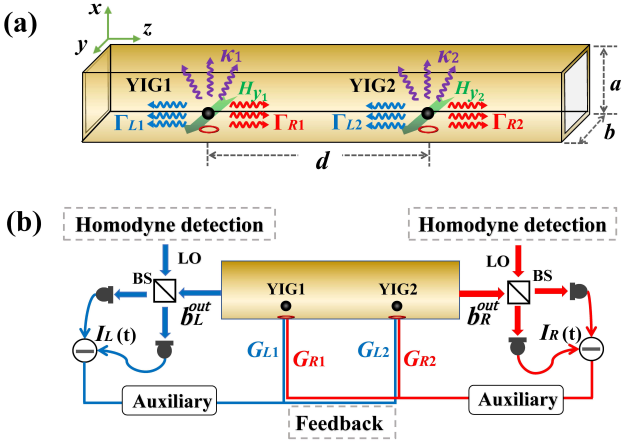}}}
\caption{(a) Chiral waveguide electromagonics. Two YIG spheres with a distance $d$ are placed in a waveguide paralleled to the $ z$ direction. The static bias magnetic field $H_{y_j}~(j=1,2)$ is along the $y$ dicrection. Superconducting microwave coils with a small loop antenna are attached to the bottom of each YIG sphere to directly drive magnon modes along the $x$ direction. Microwave photons are emitted from each sphere into the guided left- and right-propagating modes of the waveguide, with asymmetric emission rates $\Gamma_{Lj}$ and $\Gamma_{Rj}$, and the spheres are also damped by other decohering enviornments with the rates $\kappa_j$. (b) Measurement-based control scheme. The waveguide's outputs $b_{\lambda}^{\rm out}~(\lambda=L,R)$ is subjected to continuous homodyne detection. Based on the detection outcomes $I_{L, R}(t)$, indirect (state-based) feedbacks with gains $G_{\lambda j}$ are employed to achieve unconditional entanglement between the macroscopic YIG spheres.}
\label{sys}
\end{figure}

\section{Chiral magnon-waveguide system}
As shown in Fig.\ref{sys} (a), we consider a chiral magnon-waveguide system. It consists of a microwave waveguide whose modes propagating along the $z$ direction and two ferrimagnetic YIG microspheres, located at the position $z_j~(j=1,2)$ with a distance $d$, are placed in the waveguide. The uniform
magnetic field $H_{y_j}$, biased along the $y$ direction to saturate the magnetization in the spheres, produces a uniform magnon mode resonates at frequency $\omega_{mj}$ =$\gamma_0 H_{y_j}$, with the gyromagnetic ratio $\gamma_0=28$ \rm{GHz/T}. To produce the magnon entanglement between two spheres, we consider magnon Kerr nonlinear effects, resulting from the magnetocrystalline anisotropy in the YIG spheres \cite{ZZ}, which has been demonstrated by recent experimental realization of field bistability and multistability in cavity electromagnonics \cite{YPW,RCS}. To excite the system,  each YIG sphere is considered to be driven along the $x$ direction, with frequency $\omega_d$, strengths $\mathcal E_j$ and drive phases $\phi_j$, by a superconducting microwave line with a small loop antenna at its bottom. We also assume the diameters of the YIG spheres are much smaller than the wavelength of waveguide photons such that the couplings of the Kittel modes to the waveguide modes are independent to the sizes of the spheres. In the rotating frame of the driving frequency, the Hamiltonian of the whole system is of the form ($\hbar=1$)
\begin{equation}
\hat H=\hat H_m+\hat H_w+\hat H_{\rm int},
\label{1}
\end{equation}
where
\begin{align}
\hat H_m&=\sum_{j=1,2}\delta_j\hat m_j^\dag\hat m_j+K_j\hat m_j^\dag\hat m_j\hat m_j^\dag\hat m_j+i\mathcal{E}_j(\hat m_j^\dag e^{i\phi_j}\nonumber\\
&~~~ -\hat m_j e^{-i\phi_j}),\nonumber\\
\hat H_{w}&=\sum_{\lambda=L,R}\int \omega\hat b_\lambda^\dag(\omega)\hat b_\lambda(\omega)d\omega,\nonumber\\
\hat H_{\rm int}&=i\sum_{\lambda=L,R}\sum_{j=1,2}\int\frac {d\omega}{\sqrt{2\pi}}\big[g_{\lambda j }\hat b_\lambda^\dag(\omega)\hat m_j e^{-i\frac{\omega}{v_\lambda} z_j-i\omega_dt}\nonumber\\
&~~~-{\rm H.c.}\big].
 \label{Hm}
\end{align}
Here the annihilation (creation) operator $\hat m_j~(\hat m_j^\dag)$ denotes the $j$th magnon modes  and  $\hat b_\lambda~(\hat b_\lambda^\dag) ~(\lambda = L,R)$ the left- and right-propagating modes with frequency $\omega$ and wave number $k_\lambda=\omega/v_\lambda$ for the group velocity $v_\lambda$. The detuning $\delta_{j}=\omega_{mj}-\omega_d$, and the Kerr nonlinearity $K_j=\mu_0 K_{\rm{an}}\gamma_0^2/M^2V_{j}$, where $K_{\rm an}$ is the first-order anisotropy constant of the YIG samples, $M$ the saturation magnetization and $V_j$ the volume of the spheres, and $\mu_0$ the vacuum permeability. The magnon-waveguide coupling $g_{\lambda j }=\mu_0\sqrt{\frac{\gamma_0MVj}{2}}(-B_{z_j}^\lambda+iB_{x_j}^\lambda)$, with $B_{z_j}^\lambda$ ($B_{x_j}^\lambda$) being the magnetic field of the waveguide modes at the position of YIG spheres. For ${\rm TE}_{10}$ mode, $g_{\lambda_j}=\sqrt{\frac{\gamma_0MV_j}{2\epsilon_0\omega ab}}[\frac{\pi}{a}{\rm cos}(\frac{\pi x_j}{a})-k_\lambda{\rm sin}(\frac{\pi x_j}{a})]$ \cite{TYu1,TYu2}, with $a$ and $b$ being the rectangular cross section $(a\geq b)$ and $\epsilon_0$ the vacuum permittivity, which depends on the wavevector $k_\lambda$ and thus can be tuned to be chiral ($g_{Lj}\neq g_{Rj}$).
Essentially, the chirality roots from the elliptically-polarized magnetic components giving rise to the so-called spin-momentum locking phenomenon \cite{JDac,PLo}.

By treating the continua of the modes of the waveguide as reservoirs of the magnon modes, with the Born-Markovian approximation \cite{CWG}, the   master equation for the density operator $\hat \rho$ of the magnons can be written as
\begin{align}\label{Sma}
  \frac{d}{dt}\hat\rho&=-i[\hat H_m,\hat \rho]\nonumber\\
  &~~~~-{\rm Tr}_w\int_0^td\tau\Big[\hat {\tilde H}_{\rm int}(t),\big[\hat {\tilde H}_{\rm int}(\tau),\hat\rho(\tau)\otimes\hat \rho_w(0)\big]\Big],
\end{align}
where $\hat {\tilde H}_{\rm int}(t)=e^{-i\hat H_{w}t}\hat H_{\rm int}(t)e^{i\hat H_{w}t}$ and $\hat \rho_w(0)$ denotes the initial states of the waveguide's modes. By assuming initial vacua for $\hat \rho_w(0)$ and tracing out the reservoir variables, the final master equation of system is derived as (see the Appendix A)
\begin{align}
\frac{d}{dt}\hat\rho&=-i[\hat H_m+\hat H_L+\hat H_R,\hat\rho]+\sum_{\lambda=L,R}\Gamma_\lambda\mathcal{L}[\hat M_\lambda]\hat\rho\nonumber\\
&~~~+\sum_{j=1,2}\kappa_{j}\big\{(\bar n_j+1)\mathcal{L}[\hat m_j]\hat\rho+\bar n_j\mathcal{L}[\hat m_j^\dag]\hat\rho\big\},
\label{drho}
\end{align}
with the notation $\mathcal{L}[\hat o]\hat\rho\equiv\hat o\hat\rho\hat o^\dag-\{\hat o^\dag\hat o,\hat \rho\}/2$. The last line describes the magnon modes are intrinsically damped with the damping rates $\kappa_j$ by thermal environments, with the mean thermal excitation numbers $\bar n_j\equiv1/(e^{\hbar\omega_{mj}/k_BT}-1)$ at temperature $T$, $k_B$ the Boltzmann constant.  The Hamiltonian
\begin{align}
\hat H_L&\equiv-\frac{i\Gamma_L}{2}(\hat m_1^\dag\hat m_2e^{ikd}-{\rm H.c.}),\\
\hat H_R&\equiv-\frac{i\Gamma_R}{2}(\hat m_2^\dag\hat m_1e^{ikd}-{\rm H.c.}),
\end{align}
describe the coherent coupling of magnons mediated by the left and right moving photons with the wave vectors $k_R=-k_L=k$, respectively. The terms related to collective operator
\begin{align}
\hat M_\lambda=\hat m_1+\hat m_2e^{-ik_\lambda d}
\end{align}
effectively describe the dissipative-driven collective dynamics of two magnons immersed in the environments, with decay rate $\Gamma_\lambda=g_\lambda^2$.
Note that in deriving the above master equation, the time delay effect is neglected by assuming that the timescale $\Gamma_\lambda^{-1}$ of the system's evolution is much larger than the photon traveling time between the two spheres.

When the decay rates $\Gamma_L=\Gamma_R=\Gamma$, Eq.(\ref{drho}) reduces to
\begin{align}
\frac{d}{dt}\hat\rho&=-i\big[\hat H_m+\sum_{j,l=1,2}\Gamma{\rm sin}(k|z_j-z_l|)\hat m_j^\dag\hat m_l,\hat\rho\big]\nonumber\\
&~~~+\sum_{j,l=1,2}2\Gamma{\rm cos}(k|z_j-z_l|)(\hat m_l\hat\rho\hat m_j^\dag-\frac{1}{2}\{\hat m_j^\dag\hat m_l,\hat\rho\})\nonumber\\
&~~~+\sum_{j=1,2}\kappa_{j}\big\{(\bar n_j+1)\mathcal{L}[\hat m_j]\hat\rho+\bar n_j\mathcal{L}[\hat m_j^\dag]\hat\rho\big\},
\label{cas1}
\end{align}
which describes balanced bidirectional coupling between the magnon modes in the spheres. When either of the decay rates, e.g., $\Gamma_L=0$, the master equation Eq. (\ref{drho}) becomes into
\begin{align}
\frac{d}{dt}\hat\rho&=-i[\hat H_m,\hat\rho]+\sum_{j=1,2}\Gamma_R\mathcal{L}[\hat m_j]\hat\rho\nonumber\\
&~~~+\Gamma_R([\hat m_2,\hat\rho\hat m_1^\dag]e^{-ikd}-[\hat m_2^\dag,\hat m_1\hat\rho]e^{ikd})\nonumber\\
&~~~+\sum_{j=1,2}\kappa_{j}\big\{(\bar n_j+1)\mathcal{L}[\hat m_j]\hat\rho+\bar n_j\mathcal{L}[\hat m_j^\dag]\hat\rho\big\},
\end{align}
which then describes the cascade coupling between the two separate magnon modes, i.e., the second magnon mode is coupled to the first one but not vice versa \cite{Kst}. Therefore, we define
\begin{align}
D=\frac{\Gamma_R-\Gamma_L}{\Gamma_R+\Gamma_L},
 \end{align}
to characterize the chirality of the system, and $0<D\le 1$. For the balanced bidirectional situation in Eq.(\ref{cas1}), the chirality $D=0$, i.e., the nonchiral case, while for the cascade coupling the chirality $D=1$, the fully chiral case.

For strong driving of magnon modes, the Hamiltonian $\hat H_m$ can be linearized by replacing the operators $\hat m_j\rightarrow\langle \hat m_j\rangle_{\rm ss}+\hat m_j$, with steady-state amplitudes of the magnon modes $\langle \hat m_j\rangle_{\rm ss}$, and just keeping the second-order terms, it is given by
\begin{align}
\hat H_{\rm lin}=\sum_{j=1,2}\Delta_{j}\hat m_j^\dag\hat m_j+\widetilde{K}_j\big(\hat m_j^2+\hat m_j^{\dag2}\big).
\end{align}
It describes a detuned magnon parametric amplifier (MPA), with the strengths $\widetilde K_j=K_j|\langle m_j\rangle_{\rm ss}|^2$ and detuning $\Delta_{j}=\delta_{j}+4K_j |\langle \hat m_j\rangle_{\rm ss}|^2$. The amplitudes

\begin{align}
\langle \hat m_1\rangle_{\rm ss}=\frac{\mathcal{E}_1e^{i\phi_1}-\Gamma_L\langle m_2\rangle_{\rm ss}e^{ikd}}{\widetilde\Gamma_1+i\Delta_1-2i\widetilde K_1},\nonumber\\
\langle \hat m_2\rangle_{\rm ss}=\frac{\mathcal{E}_2e^{i\phi_2}-\Gamma_R\langle m_1\rangle_{\rm ss}e^{ikd}}{\widetilde\Gamma_2+i\Delta_2-2i\widetilde K_2},
\end{align}
with $\widetilde\Gamma_j=(\kappa_j+\Gamma_L+\Gamma_R)/2$.
Specifically, when $\widetilde\Gamma_j=\widetilde\Gamma$ and $\Delta_j=0$, the symmetric MPAs with
\begin{align}
\widetilde  K_1=\widetilde  K_2\approx\frac{(K_1\sqrt{K_2}\mathcal{E}_1e^{i\phi_1}-K_2\sqrt{K_1}\mathcal{E}_2e^{i\phi_2})^2}{(K_1\Gamma_L-K_2\Gamma_R)^2e^{2ikd}}
\end{align}
and the asymmetric MPAs with
\begin{align}
\widetilde  K_1=\frac{i(\Gamma_R\mathcal{E}_1e^{i(\phi_1+kd)}-\widetilde\Gamma\mathcal{E}_2^{i\phi_2})}{2\mathcal{E}_2^{i\phi_2}}~~{\rm and}~~\widetilde  K_2=0
\end{align}
can be achieved. In both cases, the strength $\widetilde K_j$ is  adjustable by changing the driving amplitudes $\mathcal E_j$.

\section{Continuous homodyne detection on waveguide's outputs}
To control the magnon systems, we consider that the waveguide's outputs  $\hat b_\lambda^{\mathrm{out}}$ is subjected by homodyne detection. For the waveguide, the input-output relation for the left and right ends reads (See the Appendix B)
\begin{equation}
\hat b_\lambda^{\mathrm{out}}(t)=\hat b_\lambda^{\mathrm{in}}(t)+\sqrt{\Gamma_\lambda}\hat M_\lambda,
  \label{bout}
\end{equation}
where $\hat b_\lambda^{\rm {in}}(t)$ is the input  vacuum noise  which satisfies the nonzero correlation $\langle\hat b_\lambda^{\rm {in}}(t)\hat b_\lambda^{\rm {in\dag}}(t')\rangle=\delta(t-t')$. We see that the outputs are related to the magnon modes and thus they can be detected by homodying the quadratures of the output fields
\begin{equation}
\hat X_{\lambda}^{\rm out}=\frac{1}{\sqrt{2}}(\hat b_\lambda^{\rm out}e^{i\theta_\lambda}+\hat b_\lambda^{\rm out\dag}e^{-i\theta_\lambda}),
  \label{X}
\end{equation}
with the local phases $\theta_\lambda$ determined by the local reference fields. The detection currents
\begin{equation}
I_{\theta_\lambda}dt=\sqrt{\eta_\lambda\Gamma_\lambda}\langle\hat M_\lambda e^{i\theta_\lambda}+\hat M_\lambda^\dag e^{-i\theta_\lambda}\rangle dt+dW_\lambda,
\label{I}
\end{equation}
where $\eta_\lambda$ is the homodyne detection efficiency and $dW_\lambda$ is the standard Wiener increments with mean zero and variance $dt$. Conditioned on the detection outcomes, the stochastic master equation for the density operator $\hat\rho_c$ is given by \cite{HMW1,JZhang}
\begin{align}
d\hat\rho_c&=-i[\hat H_{\rm lin}+\hat H_L+\hat H_R,\hat\rho_c]dt+\sum_{\lambda=L,R}\Gamma_\lambda\mathcal{L}[\hat M_\lambda]\hat\rho_cdt\nonumber\\
&~~~+\sum_{j=1,2}\kappa_{j}\big\{(\bar n_j+1)\mathcal{L}[\hat m_j]\hat\rho_cdt+\bar n_j\mathcal{L}[\hat m_j^\dag]\hat\rho_cdt\big\}\nonumber\\
&~~~+\sum_{\lambda=L,R}\sqrt{\frac{\eta_\lambda\Gamma_\lambda}{2}}\mathcal{H}[\hat M_\lambda e^{i\theta_\lambda}]\hat\rho_cdW_\lambda,
\label{drhoc}
\end{align}
with the symbols  $\mathcal H[\hat o]\hat\rho=\hat o\hat \rho+\hat\rho\hat o^\dag-\langle \hat o+\hat o^\dag\rangle$. The last term characterizes the backaction effect originating from continuously monitoring the waveguide's outputs, dependent on the measurement efficiency $\eta_\lambda$. It can be seen that for the case of full chirality, e.g., $\Gamma_L=0$,  the left output carries no information about the magnons and thus the homodyne detection on the left has null effect on the magnon system.

For the Gaussian nature of initial states, the state of the magnonic system controlled by Eq.(\ref{drhoc}) is still in Gaussian states determined by the covariance matrix $\sigma_{c,ii'}=\langle\mu_i\mu_{i'}+\mu_{i'}\mu_i\rangle/2-\langle\mu_i\rangle\langle\mu_{i'}\rangle$, where $\mu=(\hat x_{1},\hat p_{1},\hat x_{2},\hat p_{2})$ for the quadrature operators $\hat x=(\hat o+\hat o^\dag)/\sqrt{2}$ and $\hat p=-i(\hat o-\hat o^\dag)/\sqrt{2}~(o=m_j)$. From Eq.(\ref{drhoc}), we have
\begin{subequations}
\begin{align}
d\bar \mu^{T}&=\mathcal{A}\bar \mu^{T}dt+(\sigma_c \mathcal{C}-\mathcal{F})dW,\label{du}\\
\frac{d\sigma_c}{dt}&=\mathcal{A}\sigma_c+\sigma_c \mathcal{A}^{T}+\mathcal{D}-(\sigma_c \mathcal{C}-\mathcal{F})(\sigma_c \mathcal{C}-\mathcal{F})^{T},
\label{dsc}
\end{align}
\label{dmu}
\end{subequations}
where $\bar \mu=\langle\mu\rangle$, the drift matrix
\begin{align}
&\mathcal{A}=\left(
  \begin{array}{cccc}
      \mathcal{A}_{1} & \mathcal{A}_L & \\
     \mathcal{A}_R & \mathcal{A}_{2}\end{array}\right),
  \mathcal{A}_j=\left(
  \begin{array}{cccc}
     -\widetilde \Gamma_{j} & \Delta_{j}-2\widetilde K_j & \\
     -\Delta_{j}-2\widetilde K_j & -\widetilde{\Gamma}_{j} \\
\end{array}
\right)\nonumber\\
&~~~~~~~~~~~~~~~~\mathcal{A}_\lambda=\left(
  \begin{array}{cccc}
      -\Gamma_\lambda{\rm cos}kd & \Gamma_\lambda{\rm sin}kd & \\
     -\Gamma_\lambda{\rm sin}kd  & -\Gamma_\lambda{\rm cos}kd \\
\end{array}
\right),
\end{align}
the diffusion matrix
\begin{align}
\mathcal{D}= \begin{pmatrix}\begin{smallmatrix}\mathcal{D}_{1} &\mathcal{D}_{12} \\ \mathcal{D}_{12}^T & \mathcal{D}_{2}\end{smallmatrix}\end{pmatrix},
\end{align}
where $\mathcal{D}_{j}=[\kappa_{j}(\bar n_j+1/2)+(\Gamma_L+\Gamma_R)/2]I$ with $I$ the 2$\times$2 identity matrix,  and $\mathcal{D}_{12}=\begin{pmatrix}\begin{smallmatrix}\mathcal{D}_{+}&  \mathcal{D}_{-} \\ - \mathcal{D}_{-}  & \mathcal{D}_{+}\end{smallmatrix}\end{pmatrix}$, with $\mathcal{D}_{+}=[(\Gamma_L+\Gamma_R){\rm cos}kd]/2$ and $\mathcal{D}_{-}=[(\Gamma_L-\Gamma_R){\rm sin}kd]/2$.
The vectors
\begin{subequations}
\begin{align}
\mathcal{C}^T&=(\mathcal{C}_1,\mathcal{C}_2,\mathcal{C}_3,\mathcal{C}_4),\\
\mathcal{F}^T&=(\mathcal{F}_1,\mathcal{F}_2,\mathcal{F}_3,\mathcal{F}_4)/\sqrt{2}.
\end{align}
\end{subequations}
with
\begin{align}
\mathcal{C}_1&=\sqrt{\eta_L\gamma_L}{\rm cos}\theta_L+\sqrt{\eta_R\gamma_R}{\rm cos}\theta_R,\nonumber\\
\mathcal{C}_2&=-\sqrt{\eta_L\gamma_L}{\rm sin}\theta_L-\sqrt{\eta_R\gamma_R}{\rm sin}\theta_R,\nonumber\\
\mathcal{C}_3&=\sqrt{\eta_L\gamma_L}{\rm cos}(kd+\theta_L)+\sqrt{\eta_R\gamma_R}{\rm cos}(kd-\theta_R),\nonumber\\
\mathcal{C}_4&=-\sqrt{\eta_L\gamma_L}{\rm sin}(kd+\theta_L)+\sqrt{\eta_R\gamma_R}{\rm sin}(kd-\theta_R),\nonumber
\end{align}
$\mathcal{F}_j=\mathcal{C}_j$. We see from Eq.(\ref{dmu}) that the first moments are related to the measurement results and thus stochastic. Nevertheless, these stochastic moments are independent of the entanglement of the Gaussian states. On the contrary, the covariance matrix $\sigma_c$ is independent of the outcomes and deterministic and it completely determines the entanglement of the system. The effect of continuous homodyne measurement is embodied by the last nonlinear term of Eq.(\ref{dsc}) (originating from the last term of Eq.(\ref{drhoc})).

The stability of the present system is  guaranteed by the fact that all the  eigenvalues (real parts) of the drift matrix $\mathcal{A}$ are negative when the continuous detection does not exist, while with the detection the stable condition is
\begin{equation}\label{S}
  \mathcal{C}{\bf x_\xi}\neq{\bf 0}~ \forall~ {\bf x}_\xi: \widetilde{\mathcal{A}}{\bf x}_\xi=\xi {\bf x}_\xi
\end{equation}
with ${\rm Re}(\xi)\geq0$ and $\widetilde{\mathcal{A}}=\mathcal{A}+\mathcal{F} \mathcal{C}^T$. The above stability condition means that  even if the unconditional correlation matrix, in the absence of the measurement, is unstable or marginally stable, the conditional correlation matrix determined by Eq.(\ref{dsc})  can still be stable.

\begin{figure}[b]
\centering
\includegraphics[bb=30 20 800 250,scale=0.45]{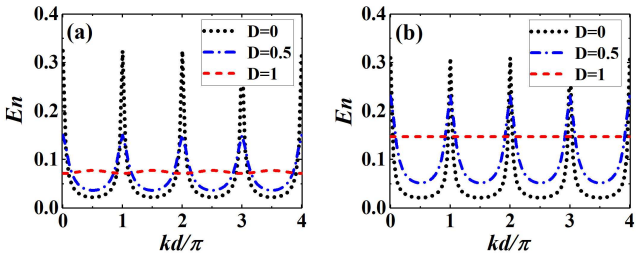}
\caption{ The magnonic entanglement $E_n$  in the steady-state regime as a function of $kd$  under different degrees of chirality, for (a) symmetric MPAs with $\widetilde K/2\pi=0.24$ {\rm MHz} and (b) asymmetric MPAs with $\widetilde K/2\pi=0.48$ {\rm MHz}, under the measurements being  absent ($\eta_L=\eta_R=0$). The other parameters are provided in the text.   }
 \label{kd}
\end{figure}

\section{Results}
We now investigate in detail the steady-state entanglement and steering between two magnon modes mediated by the waveguide.  When the covariance matrix $\sigma$ of the  system is expressed as  $\sigma=\begin{pmatrix}\begin{smallmatrix}\sigma_{1} & \sigma_{12}\\ \sigma_{12}^{T}& \sigma_{2}\end{smallmatrix}\end{pmatrix}$, the entanglement can be quantified by the logarithmic negativity $E_n$ \cite{MBP}, which is defined as
\begin{equation}\label{en}
E_{n}={\rm max}[0,-{\rm ln} (2e)],
\end{equation}
where $e=2^{-1/2}\sqrt{\Sigma-\sqrt{\Sigma^2-4{\rm det} \sigma}}$ and $\Sigma={\rm det}\sigma_{1}+{\rm det}\sigma_{2}-2{\rm det}\sigma_{12}$. From Eq.(\ref{en}), the Gaussian state is entangled if and only if $e<1/2$, which is equivalent to  Simon's necessary and sufficient entanglement nonpositive partial transpose criterion for all bipartite Gaussian states \cite{RSi}. Further, when two magnons are entangled, one intriguing property is that one magnon may steer the quantum state of the other by local operations within its own Hilbert space and by classical communication (LOCC), i.e. so-called quantum steering. To quantify the strength of steering, Kogias et al. \cite{IK}  proposed a computable measure valid for arbitrary bipartite Gaussian states based on their covariance matrix. Thus, the steering between two magnons in two directions is given by
\begin{align}\label{st}
 S_{2|1}={\rm max}[0,\frac{1}{2}{\rm ln}\frac{{\rm det}\sigma_1 }{{4\rm det}\sigma}]\\
 S_{1|2}={\rm max}[0,\frac{1}{2}{\rm ln}\frac{{\rm det}\sigma_2 }{{4\rm det}\sigma}]
\end{align}
$S_{2|1}>0$ ($S_{1|2}>0$) demonstrates that the magnonic state is steerable from the  first (second) magnon to the second (first) one. One-way steering occurs when only $S_{2|1}=0$ or $S_{1|2}=0$ holds.

\begin{figure}[t]
\centering
\includegraphics[bb=40 0 800 480,scale=0.45]{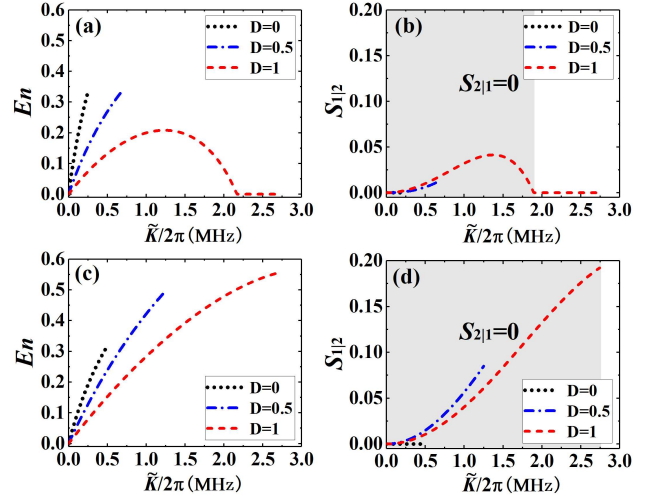}

\caption{The steady-state entanglement $E_{n}$ and steering $S_{1|2}$ vary with $\widetilde K$ under different degrees of chirality when $kd=s\pi$, for (top) symmetric and (bottom) asymmetric MPAs.
The grey areas in (b) and (d) correspond to the regions where one-way steering occurs. The related parameters are the same as Fig. \ref{kd}. In the plots, the reverse steering $S_{2|1}$ is absent and not plotted. }
 \label{ES1}
\end{figure}
\begin{figure}[t]
\centering
\includegraphics[bb=40 0 800 480,scale=0.45]{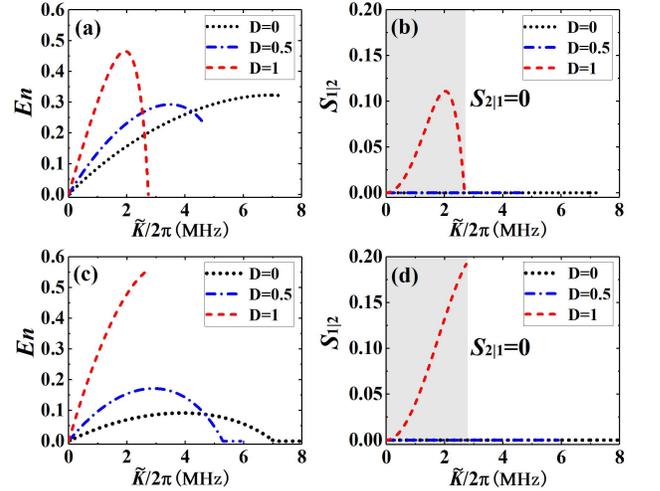}
\caption{The steady-state entanglement $E_{n}$ and steering $S_{1|2}$ vary with $\widetilde K$ under different degrees of chirality when $kd=(s+1/2)\pi$. The other settings are the same as  Fig. \ref{ES1}.}
 \label{ES2}
\end{figure}

We first consider the entanglement and steering in the absence of the measurements ($\eta_L=\eta_R=0$) for symmetric and asymmetric MPAs, i.e., $\widetilde K_1=\widetilde K_2=\widetilde K$ and $\widetilde K_1=\widetilde K$ and $\widetilde K_2=0$.  The other parameters are given by $\omega_{mj}/2\pi=10$ {\rm GHz}, $\Delta_j=0$, $\Gamma_R/2\pi=10$ {\rm MHz}, $\kappa_{j}/2\pi=1$ {\rm MHz},  $T=30$ {\rm mK} at which $\bar n_j\approx 0$. The dependence of the entanglement on the distance $d$ is plotted in Fig.\ref{kd} for different chirality degrees of $D$. It shows that the entanglement appears periodically with $kd$. In fact, the entanglement generation is due to the combination of the MPAs and coherent and dissipative couplings of the magnon modes in Eq.(\ref{drho}) which depend on the phase $kd$. When the chirality $D=0$, the maximal entanglement occurs for $kd=s\pi$ at which the coherent coupling disappears, with $s$ being an integer, whereas it becomes minimal when $kd=(s+1/2)\pi$ at which the coherent coupling exists, since the dissipative magnon coupling is more efficient than the coherent coupling for the steady-state entanglement generation. The minimal entanglement is increased while the maximal entanglement is decreased as the chirality arises, since the dissipative mixing is weaken with the increasing of the chirality. Thus, the oscillation of entanglement almost ceases with full chirality.

The dependence of the steady-state entanglement and steering on $\widetilde K_j$ is plotted in Fig.\ref{ES1} with $kd=s\pi$. As expected, the entanglement increases as $\widetilde K_j$ arises in the steady-state regime. The stability conditions
\begin{equation}
 \widetilde K<\frac{\kappa}{4}+\frac{(1-\sqrt{1-D^2})\Gamma_R}{2(1+D)},
\label{c1}
 \end{equation}
for symmetric MPAs $\widetilde K_j=\widetilde K$, and
 \begin{equation}
 \widetilde K<\frac{\kappa}{4}+\frac{\Gamma_R}{2(1+D)}-\frac{(1-D)\Gamma_R^2}{\kappa(1+D)+2\Gamma_R},
 \label{c2}
 \end{equation}
for asymmetric case $\widetilde K_1=\widetilde K$ and $\widetilde K_2=0$. We see from Eq.(\ref{c1}) that for the chirality $D=0$, the stability just depends on the non-radiation damping rate $\kappa$. This is because that for the balanced bidirectional coupling with $kd=s\pi$, a dark mode of the two magnon modes is generated and thus the stability of the whole system is determined by the dark-mode MPA with the dissipation rate $\kappa$ and independent of the radiation damping rate $\Gamma$, which in turn limits the value of the MPA strength $\widetilde K$.  When $D=1$, inequalities (\ref{c1}) and (\ref{c2}) are identical since the stability is determined by the subsystem of the first magnon mode with the cascade coupling. Thus, larger $\widetilde K$ is allowed for achieving steady states as the chirality $D$ arises for given $\Gamma_R$ and $\kappa$, as shown in Fig.\ref{ES1}. For asymmetric MPAs, this leads to the increasing of maximal achievable entanglement occurring on the thresholds, as the chirality increases. While for symmetric MPAs, the maximal entanglement decreases with full chirality, since the squeezing produced in the second MPA blocks the entanglement generation. We see that the steering is absent with non-chirality for both cases of MPAs. However, one-way steering from the second magnon mode to the first one appears when the chirality is present, as shown in Fig.\ref{ES1} (b) and (d). This means that the chirality can be used for manipulating the asymmetric steerable correlations between the magnon modes.
\begin{figure}[t]
\centering
\centering
\includegraphics[bb=20 0 800 480,scale=0.45]{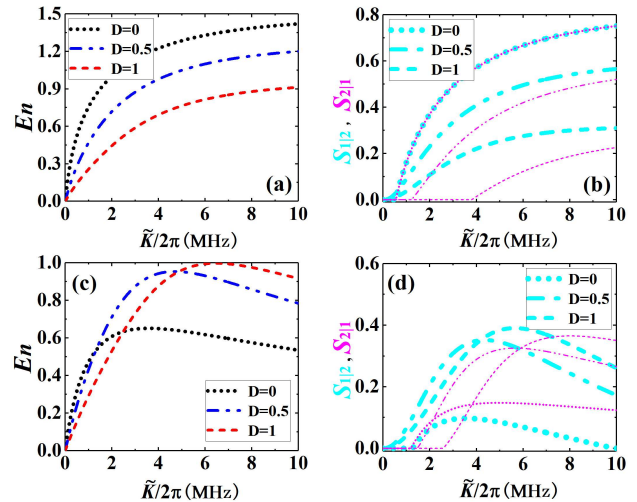}
\caption{The conditional  entanglement $E_{n}$,  steering $S_{1|2}$ (cyan thick lines) and $S_{2|1}$ (magenta thin lines) vary with $\widetilde K$ under different degrees of chirality $D$ when $kd=s\pi$, for (top) symmetric and (bottom) asymmetric MPAs, with the presence of homodyne detections ($\eta_L=\eta_R=1$). The other parameters are the same as in Fig. \ref{ES1}.  }
\label{MES1}
\end{figure}
\begin{figure}[t]
\centering
\centering
\includegraphics[bb=20 0 800 480,scale=0.45]{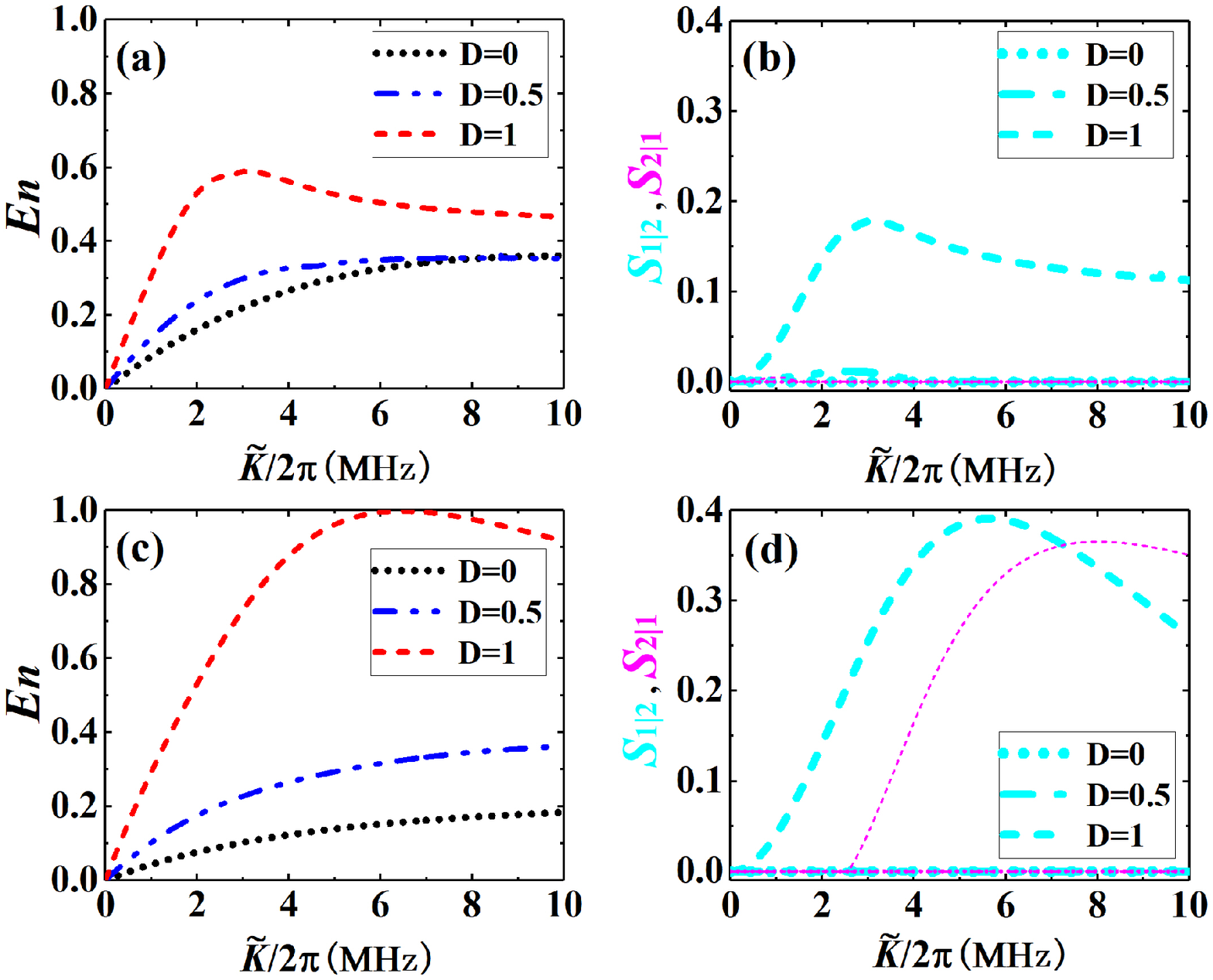}
\caption{The conditional  entanglement $E_{n}$,  steering $S_{1|2}$ (red thick lines) and $S_{2|1}$ (blue thin lines) vary with $\widetilde K$ under different degrees of chirality $D$ when $kd=(s+1/2)\pi$, for (top) symmetric and (bottom) asymmetric MPAs, with the presence of homodyne detections ($\eta_L=\eta_R=1$). The other parameters are the same as in Fig. \ref{ES2}.   }
 \label{MES2}
\end{figure}
When the distance satisfies $kd=(s+1/2)\pi$, the stability condition
\begin{equation}
 \widetilde K<\frac{\sqrt{[\kappa(1+D)+2\Gamma_R]^2+4(1-D^2)\Gamma_R^2}}{4(1+D)},
 \end{equation}
for symmetric MPAs, and
 \begin{equation}
 \widetilde K<\frac{\kappa}{4}+\frac{\Gamma_R}{2(1+D)}+\frac{(1-D)\Gamma_R^2}{\kappa(1+D)+2\Gamma_R},
 \end{equation}
and for asymmetric MPAs. In contrast with the case of $kd=s\pi$, the chirality decreases the stability regions over the MPA strength $\widetilde K$. Nevertheless, the maximal achievable entanglement also increases with the increasing of $D$, similar to that in Fig.\ref{ES1}. Moreover, it is shown that the steering is absent with non-chirality, but  is also present in the one way from the first magnon mode to the second as the chirality occurs, as already shown in Fig.\ref{ES1}. Likewise, the stronger one-way steering can be obtained for asymmetric MPAs than the symmetric case. Therefore, asymmetric MPAs setting is more favorable to the one-way steering of two magnon modes in the present system.

\begin{figure}
\centering
\includegraphics[bb=20 10 800 420,scale=0.43]{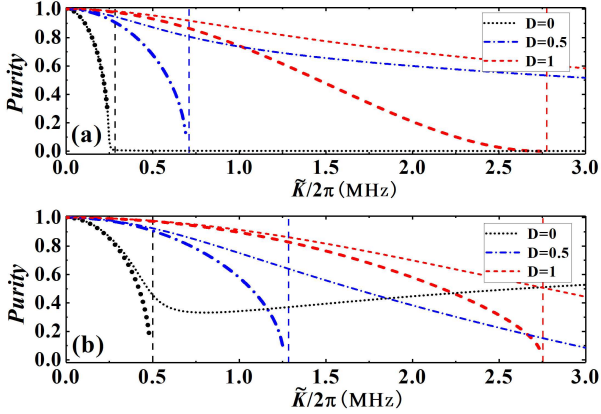}
\caption{ The purity  of the magnonic states in (a)symmetric and (b) asymmetric cases, for the measurements being absent ($\eta_L=\eta_R=0$, thick lines) and present ($\eta_L=\eta_R=1$, thin lines). The other parameters as those in Fig. \ref{MES1}.}
 \label{pur}
\end{figure}

We next study the steady-state entanglement and steering in the presence of the continuous measurement. We consider that the amplitude quadrature ($\theta_R=0$) of the right output field and phase quadrature ($\theta_L=\pi/2$) of the left output field are simultaneously subjected to homodyne detection. It should be noted that with full chirality, the detection on the left has no effect on the system. The entanglement and steering are plotted in Fig.\ref{MES1} and Fig.\ref{MES2} respectively for the cases of $kd=s\pi$ and $kd=(s+1/2)\pi$.
One can see that, compared to those in Fig.\ref{ES1} and Fig.\ref{ES2}, the entanglement and steering are considerably enhanced by the measurements no matter whether the chirality is present or not. Moreover, the reverse steering from the first magnon mode to the second is also present in the presence of the measurement. The enhancement is due to the fact that the measurement enlarges the stability region over the MPA strength and larger values of $\widetilde K$ can be allowed for achieving the steady states. Thus, the maximal achievable entanglement and steering are boosted by the measurement. On the other hand, the measurements also suppress the decoherence from the coupling of the magnons to the continua of the waveguide modes, giving rise to the enhancement of the entanglement and steering even for the same MPA strength $\widetilde K$ given in Fig.\ref{ES1} and Fig.\ref{MES1} (Fig.\ref{ES2} and Fig.\ref{MES2}). This can be partially verified by the purity of the two-mode magnon states plotted in Fig. \ref{pur} which shows that the purity is obviously enhanced by the measurement.
For symmetric MPAs, asymmetric steerings and even one-way steering can also be obtained via tuning the chirality, as demonstrated in Fig.\ref{ES1} (b) and Fig.\ref{ES2} (b). Therefore, with the measurement asymmetric steering with stronger entanglement can be achieved.

\section{Indirect feedback}
As discussed above, although the correlation matrix in Eq.(\ref{dsc}) is deterministic, the first moments $\bar\mu(t)$ depends on the detection outcomes and thus are stochastic. When an ensemble average is performed over many experimental runs, incoherent noise resulting from the random walk in phase space will mask the conditional magnon entanglement and steering. Therefore, one needs to convert the conditional results into the unconditional ones, which can be realized by introducing state-based (indirect) feedback \cite{HMW1,ACD}, different from the direct feedback in which the detection current is directly fed back to drive the system \cite{ARR}. Once the measurements are performed at some time, the values $\bar x_j(t)$ and $\bar p_j(t)$ can be inferred immediately, based on which the Markovian feedback described by the Hamiltonian
\begin{equation}
 \hat H_{\rm fb}=\sum_{\lambda=L,R}\sum_{j=1,2}G_{\lambda j}^p\bar p_{j}(t)\hat x_j-G_{\lambda j}^x\bar x_j(t)\hat p_j,
\label{hfb}
\end{equation}
can be devised, with the feedback gain parameters $G_{\lambda j}^{x,p}$. The feedback leads Eq. (\ref{du}) to be modified by substituting $\mathcal{A}$ with $\bar {\mathcal{A}}=\mathcal{A}-{\rm diag}(G_{L1}^x+G_{R1}^x,G_{L1}^p+G_{R1}^p,G_{L2}^x+G_{R2}^x,G_{L2}^p+G_{R2}^p)$.
Then, the ensemble average $\bar\sigma_e\equiv\frac{1}{2}\langle\bar\mu_i(t)\bar\mu_{i'}(t)+\bar\mu_{i'}(t)\bar\mu_{i}(t)\rangle_e$ over many realizations of the system can be derived as
\begin{equation}
\frac{d}{dt}\bar\sigma_e=\bar{\mathcal{A}}\bar\sigma_e+\bar\sigma_e\bar{\mathcal{A}}^T+(\sigma_c\mathcal{C}-\mathcal{F})(\sigma_c\mathcal{C}-\mathcal{F})^T,
\end{equation}
and  the ensemble average $\sigma_e\equiv\frac{1}{2}\langle\langle\mu_i(t)\mu_{i'}(t)+\mu_{i'}(t)\mu_{i}(t)\rangle\rangle$ is given by
\begin{equation}
  \sigma_e=\sigma_c+\bar\sigma_e,
\end{equation}
determing the system's properties under the feedback. When $\bar\sigma_e\approx0$ through choosing the appropriate feedback gains, the correlation matrix $\sigma_e\approx\sigma_c$, independent of the measurement results and thus deterministic. The overlap between the states with the covariance matrices $\sigma_c$ and $\sigma_e$ can be quantified by the fidelity \cite{LBan}
\begin{figure}[t]
\centering
\includegraphics[bb=40 0 1000 485,scale=0.46]{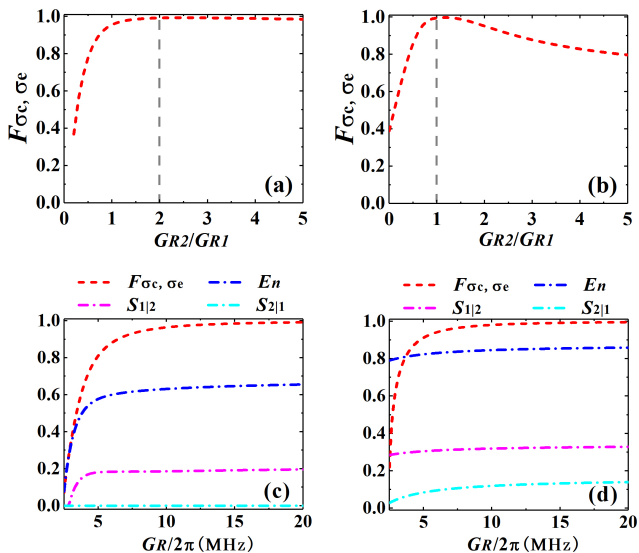}
\caption{ The effect of feedback gain parameters $G_{R2}/G_{R1}$ on fidelity $F_{\sigma_c,\sigma_e}$ for a fixed $G_{R1}/2\pi=20 {\rm MHz}$, in (a) symmetric and (b) asymmetric MPAs. (c) and (d) The dependencies of the fidelity $F_{\sigma_c,\sigma_e}$, the unconditional magnon-magnon entanglement $E_{n}$ and steerings $S_{1|2}$ and $S_{2|1}$ on feedback strengh $G_{R1}$, corresponding to the optimal feedback parameter ratios $G_{R2}/G_{R1}=2$ and $G_{R2}/G_{R1}=1$, respectively. We have chosen $D=1$, and the other parameters are the same as Fig. \ref{MES1}.   }
\label{FF}
\end{figure}
\begin{equation}
 F_{\sigma_c,\sigma_e}=(\sqrt{\Theta}+\sqrt{\Lambda}-\sqrt{(\sqrt{\Theta}+\sqrt{\Lambda})^2}-\Delta)^{-1},
\end{equation}
where $\sqrt{\Theta}=2^4{\rm det}(\Omega\sigma_c\Omega\sigma_e-\mathbbm {1}/4)$, $\Lambda=2^4{\rm det}(\sigma_c+i\Omega/2){\rm det}(\sigma_e+i\Omega/2)$, $\Delta={\rm det}(\sigma_c+\sigma_e)$, with $ \Omega= \begin{pmatrix}\begin{smallmatrix}0 & 1 \\-1 & 0\end{smallmatrix}\end{pmatrix}\otimes\mathbbm {1}$.

We take the case of the chirality $D=1$ as an example to plot the fidelity $F_{\sigma_c,\sigma_e}$ between the unconditional and conditional states of the two-mode magnon states in Fig. \ref{FF}. Since in this case only output field on the right is detected, we just consider the feedback gains $G_{L1}^{x,p}=G_{L2}^{x,p}=0$, $ G_{R1}^x=G_{R1}^p=G_{R1}$, and $G_{R2}^x=G_{R2}^p=G_{R2}$. We see that the fidelity increases as the feedback strength $G_{R1}$ arises. This is because that the increase of the feedback strength leads to stronger damping for the mean values $\bar x_j $ and $\bar p_j$, which in turn further suppress the fluctuations (i.e., $\bar \sigma_e$) of the mean values and even almost remove them completely in the limit of strong feedback.  In this limit, for the symmetric MPAs, the fidelity $F_{\sigma_c,\sigma_e}\approx 0.992$ and the entanglement and steering recover to the conditional values, $E_{n}\approx 0.656$,
$S_{1|2}\approx0.196$, and $S_{2|1}=0$ in Fig. \ref{FF}(c),  and for the asymmetric MPAs, $F_{\sigma_c,\sigma_e}\approx 0.995$, $E_{n}\approx0.859$, $S_{1|2}\approx0.328$, and $S_{2|1}\approx0.147$ in Fig. \ref{FF}(d).

\section{Discussion and Conclusion}
Before concluding, let us briefly discuss the experimental feasibility of our scheme. Firstly, a rectangular waveguide in which only the lowest ${\rm TE_{10}}$ mode exists is preferred.  The magnetic field of  ${\rm TE_{10}}$ photons is polarization-momentum locked. By changing the position of the spheres in the x-direction $x_j$, the magnon-photon coupling can be tuned to be full-chiral, partial-chiral, or non-chiral,  with achievable dissipation rates $\Gamma_{L,R}/2\pi\in(0,20)$ {\rm MHz} \cite{TYu2}. Besides,  YIG spheres with appropriate size should be selected for two reasons. The first one is to ensure the validity of the Kittel mode description for magnetic materials, i.e., the magnon excitation number should satisfy: $\langle\hat m_j^\dag\hat m_j\rangle\ll 2N_js=5N_j$; and the second one is to obtain the Kerr nonlinearity inversely proportional to the volume $V_j$. So, two submillimeter-sized YIG spheres may be ideal candidates. Moreover, to enhance the nonlinear effect, the pumping field is designed to directly drive the YIG spheres by using a superconducting MW line with a small loop antenna \cite{YPW}.  For example, considering using two YIG spheres with diameters $d_1=0.1$ {\rm mm} and $d_2=0.2$ {\rm mm}, which produce  Kerr coefficients $K_1/2\pi\approx0.0295\times10^{-6}$ {\rm Hz} and $K_2/2\pi\approx0.0132\times10^{-6}$ {\rm Hz}  respectively \cite{YPW2}. In the situation of full chirality, the symmetric MPAs setting with strength $\widetilde K_{1,2}/2\pi \approx 3$ {\rm MHz} can be achieved by the drive powers $P_1\approx0.144$ {\rm W} and $P_2\approx0.186$ {\rm W}, while the asymmetric MPAs setting with strength $\widetilde K_1/2\pi \approx 3$ {\rm MHz} and $\widetilde K_2/2\pi \approx 0$ requires drive powers  $P_1\approx0.305$ {\rm W} and $P_2\approx0.021$ {\rm W}. Proposed as in Fig. \ref{1} is a possible experimental setup design that could realize our proposal. What needs to note is that the magnetic fields of the waveguide, the driving  fields, and the uniform magnetic fields  should be orthogonal to each other at the site of the YIG spheres so that avoiding the mutual impact among them.
Furthermore, as for the verification of the quantum entanglement and steering, the method widely used in the field of cavity optomechanics can be adopted \cite{Dvi}.  Here, to read the magnon entanglement and steering, we can weakly couple each magnon mode to an independent microwave cavity acting as a probe field \cite{Jie}. Then, the magnon entangled state is transferred to the probing fields and thus the entangled state can be read out by homodyning the outputs of the probes.

In summary,  we investigate in detail quantum steerable correlations between two distant YIG spheres in a chiral microwave waveguide.  We show that for two magnons coupled to the waveguide separated by $s/2$ or $(s/2+1/4)$ wavelengths,  one-way steering can be generated using chiral magnon-photon interaction. We also find that the generated quantum steering can be enhanced considerably when the outputs of waveguide are subjected to time-continuous homodyne detection, and in this situation, the asymmetric steering with strong entanglement also can be tuned by the chirality of waveguide. To verify and apply the generated steering, we also employ optimal state-basted feedback to convert the conditional results into unconditional ones with high fidelity. Our results demonstrate the potential applicability of chirality for manipulating asymmetric steering and even one-way quantum steering. Compared to other schemes for achieving asymmetric steering,  our scheme, merely depending on the chirality enabled via positioning the micromagnets in the waveguide, is experimentally more feasible.

\section*{Acknowledgment}
This work is supported by the National Natural Science Foundation of China (Grants No.11674120 and No.12174140), the Fundamental Research Funds for the Central Universities (Grant No. CCNU20TD003) and the Excellent Doctoral Dissertation Cultivation Grant from Central China Normal University (CCNU) (Grant No. 2022YBZZ044).

\begin{widetext}
\section*{APPENDIX A: DERIVATION OF THE GENERAL CHIRAL MASTER EQUATION}
\setcounter{equation}{0}
\renewcommand{\theequation}{A.\arabic{equation}}
Here we show how to derive a general master equation for a chain of magnons coupled to a chiral waveguide. We take the general reservoir theory in quantum optics and treat the collection of magnons as the system $S$ and the bosonic modes in the chiral waveguide as a long one-dimensional reservoir $R$ exhibiting Markovian dynamics. In a rotating frame with respect to the bath Hamiltonian, the total Hamiltonian  reads
\begin{equation}\label{HH}
\hat H=\hat H_S+\hat H_{\rm int}(t),
\end{equation}
where
\begin{equation}\label{HHint}
\hat H_{\rm int}(t)=i\sum_{\lambda=L,R}\sum_j\int\frac{d\omega}{\sqrt{2\pi}}(g_{\lambda j}\hat b_\lambda^\dag(\omega)\hat m_j(t)e^{i(\omega-\omega_d)t-i\frac{\omega}{v_\lambda}z_j}-{\rm H.c.})
\end{equation}
Thus, we can get the master equation of system $\rho_S$ by tracing  out the reservoir degrees of freedom and making the Markov approximation as

\begin{align}\label{Sma}
  \frac{d\rho_S}{dt}&=-i[\hat H_S,\rho_S(t)]-i{\rm Tr}_R[\hat H_{\rm int}(t),\rho_S(0)\otimes\rho_R(0)]-{\rm Tr}_R\int_0^td\tau[\hat H_{\rm int}(t),[\hat H_{\rm int}(\tau),\rho_S(\tau)\otimes\rho_R(0)]]
\end{align}
On inserting the interaction energy Eq.(\ref{HHint}) into Eq.(\ref{Sma}), we finds

\begin{align}\label{drhos}
  \frac{d\rho_S}{dt}&=-i[\hat H_S,\rho_S(t)]+\sum_{\lambda=L,R}\sum_j\int\frac{g_{\lambda j}}{\sqrt{2\pi}}d\omega(\langle\hat b_\lambda^\dag(\omega)\rangle\hat [m_j(t),\rho_S(0)]e^{i(\omega-\omega_d)t-i\frac{\omega}{v_\lambda}z_j}-{\rm H.c.})\nonumber\\
 &~~ +\sum_{\lambda=L,R}\sum_{j,l}\int_0^td\tau\int\int\frac{g_{\lambda j}g_{\lambda l}}{2\pi}d\omega d\omega'\times\nonumber\\
 &~~\{\langle\hat b_\lambda^\dag(\omega)\hat b_\lambda^\dag(\omega')\rangle(\hat m_j(t)\hat m_l(\tau)\rho_S(\tau)-\hat m_l(\tau)\rho_S(\tau)\hat m_j(t))e^{i(\omega-\omega_d)t+i(\omega'-\omega_d)\tau-i\frac{\omega}{v_\lambda}z_j-i\frac{\omega'}{v_\lambda}z_l}\nonumber\\
 &~~+\langle\hat b_\lambda(\omega)\hat b_\lambda(\omega')\rangle(\hat m_j^\dag(t)\hat m_l^\dag(\tau)\rho_S(\tau)-\hat m_l^\dag(\tau)\rho_S(\tau)\hat m_j^\dag(t))e^{-i(\omega-\omega_d)t-i(\omega'-\omega_d)\tau+i\frac{\omega}{v_\lambda}z_j+i\frac{\omega'}{v_\lambda}z_l}\nonumber\\
 &~~-\langle\hat b_\lambda^\dag(\omega)\hat b_\lambda(\omega')\rangle(\hat m_j(t)\hat m_l^\dag(\tau)\rho_S(\tau)-\hat m_l^\dag(\tau)\rho_S(\tau)\hat m_j(t))e^{i(\omega-\omega_d)t+i(\omega'-\omega_d)\tau-i\frac{\omega}{v_\lambda}z_j+i\frac{\omega'}{v_\lambda}z_l}\nonumber\\
 &~~-\langle\hat b_\lambda(\omega)\hat b_\lambda^\dag(\omega')\rangle(\hat m_j^\dag(t)\hat m_l(\tau)\rho_S(\tau)-\hat m_l(\tau)\rho_S(\tau)\hat m_j^\dag(t))e^{-i(\omega-\omega_d)t+i(\omega'-\omega_d)\tau+i\frac{\omega}{v_\lambda}z_j-i\frac{\omega'}{v_\lambda}z_l}-{\rm H.c.}\}
\end{align}
where the expectation values refer to the initial state of the reservoir. For example, we assume that  the waveguide initially in the vacuum state $\rho_R(0)=|{\rm vac}\rangle\langle {\rm vac}|$, we have

\begin{align}\label{bb}
  \langle \hat b_\lambda(\omega)\rangle&=0,~~ \langle \hat b_\lambda^\dag(\omega)\rangle=0,\nonumber \\
  \langle \hat b_\lambda(\omega)\hat b_\lambda(\omega')\rangle&=0,~~\langle \hat b_\lambda^\dag(\omega)\hat b_\lambda^\dag(\omega')\rangle=0,\nonumber\\
  \langle \hat b_\lambda^\dag(\omega)\hat b_\lambda(\omega')\rangle&=0,~~\langle \hat b_\lambda(\omega)\hat b_\lambda^\dag(\omega')\rangle=\delta_{\omega\omega'},
\end{align}
By substituting Eq.(\ref{bb}) into Eq.(\ref{drhos}) and introducing $k_\lambda\equiv\omega_d/v_\lambda$ and $\Gamma_\lambda=g_\lambda^2$, one obtains the master equation for the evolution of the magnon chain in chiral waveguide as

\begin{align}\label{ddr}
  \frac{d\rho_S}{dt}&=-i[\hat H_S,\rho_S(t)]+\sum_{\lambda=L,R}\sum_{j,l}\sqrt{\Gamma_{\lambda j}\Gamma_{\lambda l}}\theta(\frac{z_j-z_l}{v_\lambda})([\hat m_j(t),\rho_S(t)\hat m_l(t)^\dag]e^{-ik_\lambda(z_j-z_l)}-[\hat m_j(t)^\dag,\hat m_l(t)\rho_S(t)]e^{ik_\lambda(z_j-z_l)})
\end{align}
where the function $\theta(x)$ is defined as: $\theta(x)=1$ when $x>0$, $\theta(x)=0$ when $x<0$ and $\theta(x)=1/2$ when $x=0$. It reflects the time ordering of  the magnons along the left and right propagation directions. Note that from Eq.(\ref{drhos}) to Eq.(\ref{ddr}), the integral over $\omega$ is extended to $\pm\infty$ according to the Weisskopf-Wigner approximation, and the retardation effects arising from a finite propagation velocity of the traveling photons are assumed to be neglected, i.e, $\hat m_l(t-\frac{z_j-z_l}{v_\lambda}) \approx\hat m_l(t)$.

\section*{APPENDIX B: DERIVATION OF THE INPUT AND OUTPUT RELATIONS OF WAVEGUIDE}
\setcounter{equation}{0}
\renewcommand{\theequation}{B.\arabic{equation}}

We start with the Heisenberg equations of motion for waveguide-bath operators $\hat b_\lambda(\omega,t)$, which is given by

\begin{align}\label{db}
  \frac{d}{dt}\hat b_\lambda(\omega,t)&=\sum_{j=1,2}\sqrt{\frac{\Gamma_\lambda}{2\pi}}\hat m_j e^{i(\omega-\omega_d)t-i\frac{\omega}{v_\lambda}z_j},
\end{align}
The formal solution to this equations depends on whether we choose to solve in terms of the input conditions at time $t=t_0$ or in terms of the output conditions at time $t=t_1$, which reads

\begin{align}\label{t0}
 \hat b_\lambda(\omega,t)=\hat b_{\lambda }(\omega,t_0)+\int_{t_0}^t\sum_{l=1,2}\sqrt{\frac{\Gamma_\lambda}{2\pi}}\hat m_l(t)e^{-i(\omega-\omega_d)s-i\frac{\omega }{v_\lambda} z_l}ds,
\end{align}
with $t>t_0$, or

\begin{align}\label{t1}
 \hat b_\lambda(\omega,t)=\hat b_{\lambda }(\omega,t_1)-\int_{t}^{t_1}\sum_{l=1,2}\sqrt{\frac{\Gamma_\lambda}{2\pi}}\hat m_l(t)e^{-i(\omega-\omega_d)s-i\frac{\omega } {v_\lambda}z_l}ds,
\end{align}
with $t<t_1$. The magnon operator obeys the Heisenberg equation

\begin{align}\label{dbt}
\frac{d}{dt}\hat m_j(t)&=-i[\hat m_j(t),\hat H_m(t)]-\frac{\kappa_j}{2}\hat m_j(t)-\sqrt{\kappa_j}\hat m_j^{\rm in}(t)\nonumber\\
~~~&-\sum_{\lambda=L,R}\sum_{l=1,2}\int d\omega \sqrt{\frac{\Gamma_\lambda}{2\pi}}[\hat m_j(t),\hat m_l^\dag(t)]\hat b_\lambda(\omega,t)e^{-i(\omega-\omega_d)t+i\frac{\omega}{v_\lambda}z_l},
\end{align}
Inserting the solutions (\ref{t0}) and (\ref{t1}) into Eq.(\ref{dbt}) respectively, one obtains

\begin{align}\label{in}
  \frac{d}{dt}\hat m_j(t)&=-i[\hat m_j(t),\hat H_m(t)]-\frac{\kappa_j}{2}\hat m_j(t)-\sqrt{\kappa_j}\hat m_j^{\rm in}(t)
 -\sum_{\lambda=L,R}\sum_{l=1,2}\sqrt{\Gamma_\lambda}[\hat m_j(t),\hat m_l^\dag(t)]\hat b_\lambda^{\rm in}(t)e^{ik_\lambda z_l}\nonumber\\
  ~~~&-\sum_{\lambda=L,R}\sum_{l=1,2}\frac{\Gamma_\lambda}{2}[\hat m_j(t),\hat m_l^\dag(t)]\hat m_l(t)
 -\Gamma_L[\hat m_j(t),\hat m_1^\dag(t)]\hat m_2(t)e^{ik_L(z_1-z_2)}-\Gamma_R[\hat m_j(t),\hat m_2^\dag(t)]\hat m_1(t)e^{ik_R(z_2-z_1)},
\end{align}
and
\begin{align}\label{out}
  \frac{d}{dt}\hat m_j(t)&=-i[\hat m_j(t),\hat H_m(t)]-\frac{\kappa_j}{2}\hat m_j(t)-\sqrt{\kappa_j}\hat m_j^{\rm in}(t)
 -\sum_{\lambda=L,R}\sum_{l=1,2}\sqrt{\Gamma_\lambda}[\hat m_j(t),\hat m_l^\dag(t)]\hat b_\lambda^{\rm out}(t)e^{ik_\lambda z_l}\nonumber\\
  ~~~&+\sum_{\lambda=L,R}\sum_{l=1,2}\frac{\Gamma_\lambda}{2}[\hat m_j(t),\hat m_l^\dag(t)]\hat m_l(t)
 +\Gamma_L[\hat m_j(t),\hat m_2^\dag(t)]\hat m_1(t)e^{ik_L(z_2-z_1)}-\Gamma_R[\hat m_j(t),\hat m_1^\dag(t)]\hat m_2(t)e^{ik_R(z_1-z_2)},
\end{align}
where we have defined the input and output fields as

\begin{align}\label{inout}
 \hat b_\lambda^{\rm in}=\frac{1}{\sqrt{2\pi}}\int d\omega\hat b_\lambda(\omega,t_0)e^{-i(\omega-\omega_d)t},\\
  \hat b_\lambda^{\rm out}=\frac{1}{\sqrt{2\pi}}\int d\omega\hat b_\lambda(\omega,t_1)e^{-i(\omega-\omega_d)t},\\
\end{align}
Therefore, by subtracting Eq.(\ref{out}) from Eq.(\ref{in}), the input-output relations for  both ends of the waveguide can be derived as Eq.(\ref{bout}) in Sec.III.

\end{widetext}

\end{document}